\documentclass[11pt]{article}

\usepackage[a4paper,margin=1in]{geometry}
\usepackage{amsmath,amssymb,amsthm,mathtools}
\usepackage{booktabs}
\usepackage[numbers,sort&compress]{natbib}
\usepackage[colorlinks=true,linkcolor=blue,citecolor=blue,urlcolor=blue]{hyperref}
\usepackage{microtype}

\newtheorem{theorem}{Theorem}
\newtheorem{lemma}[theorem]{Lemma}
\newtheorem{proposition}[theorem]{Proposition}

\theoremstyle{remark}

\newcommand{\F}{\mathbb{F}_2}
\newcommand{\E}{\mathbb{E}}
\newcommand{\Prob}{\mathbb{P}}
\newcommand{\nul}{\operatorname{nul}}
\newcommand{\rank}{\operatorname{rank}}
\newcommand{\wt}{\operatorname{wt}}
\newcommand{\supp}{\operatorname{supp}}
\newcommand{\Rnd}{\mathcal{R}_{n,d}}
\newcommand{\Bnd}{\mathcal{B}_{n,d}}
\newcommand{\NA}{N_{\mathrm{NA}}}
\newcommand{\NAd}{N_{\mathrm{A}}}

\title{Learning Sparse Graph States with Local Pauli Measurements:\\
Tight Nonadaptive Bounds and an Adaptive Separation}
\author{Matteo De Francesco\\
University of Geneva, Switzerland\\
\texttt{matteo.defrancesco@unige.ch}}
\date{}

\begin{document}
\maketitle

\begin{abstract}
We study how many copies of an unknown $n$-qubit graph state are needed to
identify its graph when every qubit is measured in the $X$, $Y$ or $Z$
basis. Ouyang and Tomamichel showed that $O(d^2\log n)$ copies suffice for
$d$-regular graphs, while any measurement needs $\Omega(d\log n)$, and
asked whether the quadratic dependence on $d$ is necessary. We show that
it is for nonadaptive learners: if $d^3\ln n\le n$, every nonadaptive
strategy needs $\Omega(d^2\log n)$ copies, even with an unrestricted
classical decoder and even if a small constant fraction of the edges may
be reported wrongly. We then show that adaptivity between copies can
beat this barrier. On connected bipartite $d$-regular graphs of
logarithmic diameter whose bipartition is hidden, nonadaptive learners
still need $\Theta(d^2\log n)$ copies, but a breadth-first adaptive learner
needs only $O(d\log^2 n/\log\log n)$; revealing the bipartition brings
the nonadaptive cost down to $\Theta(d\log n)$. The lower bounds rest on a
simple fact: one copy reveals at most the binary nullity of a principal
submatrix of the adjacency matrix. We control this nullity for random
unions of perfect matchings through Fourier estimates and, in the
bipartite case, a method-of-types argument.
\end{abstract}


\section{Introduction}

Graph states are the stabilizer states naturally attached to graphs. They
are the resource of measurement-based quantum computation and appear
throughout quantum error correction and quantum networks
\citep{hein2004}. A device that is supposed to prepare a given graph state
may prepare a different one, and it is natural to ask how much data is
needed to find out which graph was actually prepared. On current hardware
the cheapest measurements are single-qubit Pauli measurements, so we ask
the question in that model: given $N$ independent copies of $|G\rangle$,
where $G$ is an unknown $d$-regular graph on $n$ labelled vertices, and
measuring each qubit of each copy in the $X$, $Y$ or $Z$ basis, how large
must $N$ be to recover $G$?

\citet{ot2022} gave a simple randomized algorithm for this task that uses
$O(d^2\log n)$ copies and only $X/Z$ measurements, and proved that
$\Omega(d\log n)$ copies are necessary even for arbitrary joint
measurements. Bell sampling on pairs of copies achieves $O(d\log n)$
\citep{ms2022}. Ouyang and Tomamichel left open whether the quadratic
dependence on $d$ is necessary for product measurements, and whether
adaptive strategies can do better.

\paragraph{Results.}
Let $\Rnd$ be the set of simple $d$-regular graphs on $[n]$. A
\emph{nonadaptive} learner fixes the Pauli settings of all copies in
advance, possibly at random and with arbitrary correlations; an
\emph{adaptive} learner may choose the setting of each new copy as a
function of all earlier outcomes. A learner succeeds if it outputs $G$
with probability at least $2/3$ for every graph in the promise class.

\begin{theorem}\label{thm:regular}
For all sufficiently large $n$ and all $d\ge 2$ with $nd$ even and
$d^3\ln n\le n$, the number of copies needed to learn graphs in $\Rnd$
with nonadaptive single-qubit Pauli measurements is $\Theta(d^2\log n)$.
\end{theorem}

The upper bound is the algorithm of \citet{ot2022}; our contribution is
the matching lower bound, which holds for every randomized schedule and
every classical decoder. For $d=\Theta(\log n)$ the answer is
$\Theta(\log^3 n)$.

The second result shows that adaptivity can help. Let
$D_n=\lceil 16\ln n/\ln\ln n\rceil$ and let $\Bnd\subseteq\Rnd$ consist of
the connected bipartite graphs of diameter at most $D_n$. The learner is
\emph{not} told the bipartition.

\begin{theorem}\label{thm:bipartite}
For all sufficiently large even $n$ and all $d$ with $\lceil\ln n\rceil\le
d$ and $d^3\ln n\le n$,
\begin{align*}
\NA(\Bnd)&=\Theta(d^2\log n)&&\text{for nonadaptive $X/Y/Z$ learners, and}\\
\NAd(\Bnd)&=O\!\left(\frac{d\log^2 n}{\log\log n}\right)&&\text{for adaptive $X/Z$ learners.}
\end{align*}
\end{theorem}

The gain is a factor $\Omega(d\log\log n/\log n)$; for $d=n^{1/4}$, for
example, the two bounds are $\Theta(n^{1/2}\log n)$ and
$O(n^{1/4}\log^2 n/\log\log n)$. The hidden bipartition is essential: if
it is revealed, a single repeated nonadaptive setting already achieves
$\Theta(d\log n)$ (Proposition~\ref{prop:revealed}). So the extra factor
$d$ is the price of not knowing which large sets of qubits are
independent, even though the bipartition itself is only $n$ bits of
information.

Finally, the lower bounds do not depend on exact recovery.

\begin{theorem}\label{thm:approx}
There is an absolute constant $\varepsilon_0>0$ such that, for every
$0<\varepsilon\le\varepsilon_0$, outputting a graph $\widehat G$ with
$|E(\widehat G)\,\triangle\,E(G)|\le\varepsilon nd$ with probability at least
$2/3$ still requires $\Omega(d^2\log n)$ nonadaptive Pauli copies on
$\Rnd$ and on $\Bnd$, in the parameter ranges of
Theorems~\ref{thm:regular} and~\ref{thm:bipartite}.
\end{theorem}

\begin{table}[t]
\centering
\small
\begin{tabular}{@{}lll@{}}
\toprule
Promise & Measurements & Copies\\
\midrule
$\Rnd$ & arbitrary & $\Theta(d\log n)$\\
$\Rnd$ & nonadaptive $X/Y/Z$ & $\Theta(d^2\log n)$ (Thm.~\ref{thm:regular})\\
$\Bnd$, hidden sides & nonadaptive $X/Y/Z$ & $\Theta(d^2\log n)$ (Thm.~\ref{thm:bipartite})\\
$\Bnd$, hidden sides & adaptive $X/Z$ & $O(d\log^2 n/\log\log n)$ (Thm.~\ref{thm:bipartite})\\
$\Bnd$, sides revealed & nonadaptive $X/Z$ & $\Theta(d\log n)$ (Prop.~\ref{prop:revealed})\\
\bottomrule
\end{tabular}
\caption{Copy complexity of learning $d$-regular graph states (constant
success probability, degree conditions as in the theorems). The first
row combines the lower bound of \citet{ot2022} with Bell sampling
\citep{ms2022}.}
\label{tab:results}
\end{table}

\subsection{Overview of the techniques}

\paragraph{What one copy can tell.}
Fix a setting and let $W$ be the set of qubits measured in $X$ or $Y$.
The outcome of one copy is uniformly distributed on an affine subspace,
cut out by one parity check for each vector in the kernel of
$A_G[W,W]+\Lambda$, where $A_G$ is the adjacency matrix over $\F$ and
$\Lambda$ marks the $Y$ qubits (Lemma~\ref{lem:outcome}). Everything else
is noise. Hence one copy carries at most $\nul(A_G[W,W]+\Lambda)$ bits
about $G$. For intuition, think of $X/Z$ settings: an $X$-measured vertex
$v$ whose neighbours are all measured in $Z$ yields the deterministic
parity of its stabilizer $X_vZ_{N(v)}$, and this is exactly the test used
by the Ouyang--Tomamichel algorithm. Two adjacent $X$-measured vertices
scramble each other. A learner who does not know the graph cannot choose
a large set $W$ without internal edges, and for a random $d$-regular graph
a typical set of size $n/d$ contains only about $n/d$ vertices with no
neighbour in $W$.

\paragraph{From one copy to a lower bound.}
For a nonadaptive learner the settings are independent of the graph, so
the information in $N$ copies is at most $N\sup_{W,\Lambda}\E\,\nul(A_G[W,W]+\Lambda)$
for any prior on $G$ (Lemma~\ref{lem:info}). A Fano-type argument shows
that about $\log(1/\max_g\pi(g))$ bits are needed (Lemma~\ref{lem:fano}).
It therefore suffices to find a distribution on $d$-regular graphs with
min-entropy $\Omega(nd\log n)$ under which every fixed setting has
expected nullity $O(n/d)$.

\paragraph{The hard distribution.}
We take $G$ to be the union of $d$ independent uniform perfect matchings,
conditioned on being disjoint. Before conditioning, independence turns
the exponential moment $\E\,2^{\nul}$ into a sum over pairs of vectors of
the $d$-th power of a single sign average, which we read as a random
pairing of vertices carrying four colours. A switching argument shows that
the sign average is tiny unless the colouring is close to two-coloured
(Lemma~\ref{lem:fourcolour}), and counting the nearly two-coloured
configurations gives $\E\,2^{\nul}\le e^{O(n/d)}$
(Proposition~\ref{prop:moment}). We work with $2^{\nul}$ rather than
$\nul$ because it has an exact Fourier expansion and because conditioning
on the rare event of disjointness then only costs an additive
$O(d^2\log n)$, which the condition $d^3\ln n\le n$ absorbs.

\paragraph{The bipartite family.}
On $\Bnd$ we need a new hard distribution: a uniformly random hidden
bipartition together with $d$ random bijections between the two sides.
For $X/Z$ settings the nullity splits into an imbalance term, which is
$O(\sqrt n)$ on average because the bipartition is hidden, and a rank
deficiency of a rectangular random matrix. To handle $Y$ measurements we
compare the shifted matrix with an unshifted one using an even number of
matchings (Lemma~\ref{lem:shift}) and bound the resulting second moment of
the rectangular kernel by the method of types (Lemma~\ref{lem:second}). An
expansion argument shows that the same random graphs have diameter
$O(\log n/\log\log n)$. The adaptive learner exploits bipartiteness
directly: every breadth-first layer is an independent set, so all its
neighbourhoods can be read off in parallel.

\subsection{Related work}

Information-theoretic lower bounds for learning phase and stabilizer
states with single-copy measurements are known for dense instances:
\citet{abdy2023} prove an $\Omega(n^2)$ bound for quadratic phase states,
\citet{as2026} extend it to limited quantum memory, and
\citet{begms2026} show that adaptive single-copy Clifford measurements,
which may entangle qubits within a copy, reduce the cost to $O(n)$. These
dense ensembles say nothing about sparse regular graphs, where the
relevant parameter is $d$ rather than $n$. General single-copy lower
bounds without graph structure are given by \citet{ln2022}, and
\citet{lto2026} analyse specific product-measurement learners for
stabilizer codes.

Our goal, exact or approximate identification of the graph, differs from
quantum PAC learning, where one predicts expectation values of test
measurements drawn from a distribution; stabilizer states are efficiently
PAC-learnable in that sense \citep{rocchetto2018}. Graph states are phase
states with a quadratic phase, so they also fit the agnostic phase-state
framework of \citet{adgo2026}, which targets fidelity guarantees for
Boolean circuit classes and uses different resources. A classical
analogue of the adaptive/nonadaptive gap appears in learning hidden
graphs with edge-detecting queries \citep{ab2019}, although an
edge-detecting query and a measurement transcript are quite different
observations.

On the probabilistic side, sums of random matchings and permutation
matrices over finite groups are studied by \citet{meszaros2020}, whose
mixing results concern full vectors rather than the principal and
rectangular marginals we need; \citet{portier2026} relates conditioned
matching unions to uniform random regular graphs in total variation,
which is too weak to transfer an $O(n/d)$ nullity bound. Random-angle
moments of regular graph states are studied by \citet{ghh2025}.

\section{Model and the information in one copy}\label{sec:model}

For a simple graph $G$ on $[n]$ with adjacency matrix $A=A_G$ over $\F$,
let $q_G(x)=\sum_{i<j}A_{ij}x_ix_j$ (the number of edges inside the
support of $x$, modulo two) and
\[
|G\rangle=2^{-n/2}\sum_{x\in\F^n}(-1)^{q_G(x)}|x\rangle .
\]
The state is stabilized by $K_v=X_v\prod_{u\in N(v)}Z_u$ for every vertex
$v$. Ranks and nullities are over $\F$, $\ln$ is the natural logarithm and
$\log$ is base two.

A \emph{setting} measures each qubit in $X$, $Y$ or $Z$. We encode it by
the set $W$ of qubits measured in $X$ or $Y$ and the $0/1$ diagonal matrix
$\Lambda$ on $W$ marking the $Y$ qubits; the qubits in $U=[n]\setminus W$
are measured in $Z$. We write $C=A[W,W]$ and $B=A[W,U]$, and record the
outcomes as bits $s\in\F^W$ and $t\in\F^U$, a bit $b$ standing for the
eigenvalue $(-1)^b$. For $a\in\F^W$ let
$\omega(a)=q_G(a)+\tfrac12\wt(\Lambda a)\bmod 2$, where $\wt$ is the
Hamming weight.

\begin{lemma}[Outcome law]\label{lem:outcome}
Let $K=\ker(C+\Lambda)$. For every $a\in K$ the weight $\wt(\Lambda a)$ is
even, $\omega$ is linear on $K$, and the outcome $(s,t)$ is uniformly
distributed on the affine subspace
\begin{equation}\label{eq:affine}
a\cdot s+a^{\mathsf T}Bt=\omega(a)\qquad (a\in K),
\end{equation}
which has dimension $n-\nul(C+\Lambda)$. In particular the outcome has
entropy $n-\nul(A[W,W]+\Lambda)$ bits.
\end{lemma}

Each $a\in K$ corresponds to a stabilizer of $|G\rangle$ that is a product of
measured observables; \eqref{eq:affine} says that its value is fixed,
and all other outcome parities are fair coins. For instance, if $v\in W$
is an $X$ qubit with no neighbour in $W$, then $a=e_v$ gives
$s_v=\sum_{u\in N(v)}t_u$. The sign $\omega$ cannot be dropped: for a
triangle measured entirely in $X$, the product of the three outcomes is
always $-1$. The proof is in Appendix~\ref{app:model}.

\begin{lemma}[Information]\label{lem:info}
Let $\pi$ be a distribution on graphs and let $S=((W_j,\Lambda_j))_{j\le N}$
be a random schedule independent of $G$. The record $Y$ of the $N$ copies
satisfies
\[
I(G;Y,S)\le N\sup_{W,\Lambda}\E_{G\sim\pi}\nul(A_G[W,W]+\Lambda).
\]
\end{lemma}

\begin{proof}
Since $S$ is independent of $G$, $I(G;Y,S)=I(G;Y\mid S)\le H(Y\mid
S)-H(Y\mid G,S)$. The first term is at most $Nn$. Given $G$ and $S$ the
copies are independent, so by Lemma~\ref{lem:outcome}
$H(Y\mid G,S)=\sum_j\E[n-\nul(A_G[W_j,W_j]+\Lambda_j)]$. Because each
$(W_j,\Lambda_j)$ is independent of $G$, each expectation is at most the
supremum over fixed settings.
\end{proof}

The independence of the schedule from $G$ is used in the last step. For
an adaptive learner the setting of a copy depends on $G$ through earlier
outcomes, and the right-hand side would become $\E\max_{W,\Lambda}\nul$,
which can be as large as $n/2$: think of a bipartite graph with $W$ one of
its sides. This is precisely the room that Theorem~\ref{thm:bipartite}
exploits.

\begin{lemma}[Identification up to a tolerance]\label{lem:fano}
Let $\pi$ be a distribution on a finite class of graphs and
$\pi_*=\max_g\pi(g)$. Suppose that every possible output is within the
allowed tolerance of at most $V$ graphs of the class, and that $V\pi_*<1$.
If an estimator computed from a record $T$ meets the tolerance with
probability at least $2/3$ when $G\sim\pi$, then
$I(G;T)\ge\tfrac23\log\frac{1}{V\pi_*}-1$.
\end{lemma}

\begin{proof}
Add the estimator's private randomness to $T$. Let $P$ be the joint law of
$(G,T)$ and $Q$ the product of its marginals, so that $I(G;T)=D(P\|Q)$.
Under $Q$ the output is independent of $G$, so the success event has
probability at most $V\pi_*$; under $P$ it has probability $r\ge 2/3$.
Data processing for the indicator of this event gives
$D(P\|Q)\ge r\log\frac1{V\pi_*}-h(r)\ge\frac23\log\frac{1}{V\pi_*}-1$,
where $h$ is the binary entropy.
\end{proof}

Exact recovery is the case $V=1$. Combining the two lemmas, a
distribution $\pi$ on the promise class gives the lower bound
\begin{equation}\label{eq:recipe}
N\;\ge\;\frac{\tfrac23\log\frac1{V\pi_*}-1}{\sup_{W,\Lambda}\E_\pi\nul(A_G[W,W]+\Lambda)} ,
\end{equation}
since a learner that succeeds on every graph of the class also succeeds
on average under $\pi$.

The upper bound in Theorem~\ref{thm:regular} is the following slight
extension of \citet{ot2022}; its proof is in Appendix~\ref{app:model}.

\begin{proposition}\label{prop:upper}
Graphs of maximum degree at most $\Delta\ge1$ can be identified with
failure probability at most $\delta$ from
$O(\Delta^2\ln(en/\Delta)+\Delta\ln(n/\delta))$ nonadaptive copies, by
measuring every qubit independently in $X$ with probability $1/(\Delta+1)$
and in $Z$ otherwise.
\end{proposition}

The two factors of $\Delta$ have different origins. A test of a candidate
neighbourhood of $v$ needs $v$ in $X$ and all its candidate neighbours in
$Z$, which happens on a fraction of order $1/\Delta$ of the copies, and a
neighbourhood carries about $\Delta\log n$ bits.

\section{The nonadaptive lower bound for regular graphs}\label{sec:regular}

\paragraph{The hard distribution.}
Let $n$ be even and let $M_1,\dots,M_d$ be independent uniformly random
perfect matchings of $[n]$. Their sum $A=A_{M_1}+\dots+A_{M_d}$ over $\F$
is the adjacency matrix of a simple $d$-regular graph exactly on the event
$\mathcal S$ that the matchings are pairwise disjoint. We let $\pi$ be the
law of the union graph conditioned on $\mathcal S$.

\paragraph{Exponential moments as sign averages.}
For any symmetric $m\times m$ matrix $C$ over $\F$,
\begin{equation}\label{eq:kernelcount}
2^{\nul C}=|\ker C|=2^{-m}\sum_{\sigma,\tau\in\F^m}(-1)^{\sigma^{\mathsf T}C\tau},
\end{equation}
because the inner sum over $\tau$ vanishes unless $C\sigma=0$. Fix $W$
with $|W|=m$ and extend $\sigma,\tau$ by zero outside $W$. Independence of
the matchings gives
\begin{equation}\label{eq:phid}
\E\,2^{\nul(A[W,W]+\Lambda)}
=2^{-m}\sum_{\sigma,\tau\in\F^W}(-1)^{\sigma^{\mathsf T}\Lambda\tau}\,\varphi(\sigma,\tau)^d,
\qquad
\varphi(\sigma,\tau)=\E_M(-1)^{\sigma^{\mathsf T}A_M\tau},
\end{equation}
where $M$ is a single uniform perfect matching. The diagonal $\Lambda$ only
contributes a fixed sign. To understand $\varphi$, give vertex $i$ the
colour $c_i=(\sigma_i,\tau_i)\in\F^2$; vertices outside $W$ get colour $0$.
Then $\sigma^{\mathsf T}A_M\tau=\sum_{\{i,j\}\in M}(\sigma_i\tau_j+\sigma_j\tau_i)$,
and the summand equals $1$ exactly when $c_i$ and $c_j$ are two different
nonzero colours. So $\varphi$ is the average, over a random pairing of the
coloured vertices, of $(-1)$ to the number of such ``bad'' pairs.

\begin{lemma}[Four-colour estimate]\label{lem:fourcolour}
Let $N_1\ge N_2\ge N_3\ge N_4$ be the sizes of the four colour classes.
For even $n\ge128$ and $\eta=2^{-29}$,
\[
|\varphi(\sigma,\tau)|\le\exp(-\eta N_2N_3/n).
\]
\end{lemma}

If only two colours occur, the parity of the number of bad pairs is
fixed and $|\varphi|=1$, so a bound of this form is the best one can hope
for. When three classes are large, the parity is close to a fair coin.
The reason is a switching structure: four vertices coloured $0,0,A,B$
admit three internal pairings, exactly one of which contains a bad pair,
so re-pairing them at random flips the parity with probability $1/3$. A
random matching contains about $N_2N_3/n$ such independent switches. The
proof (Appendix~\ref{app:regular}) makes this precise by generating the
matching in blocks of four vertices.

\begin{lemma}[Counting colourings]\label{lem:count}
For every $W$ with $|W|=m$ and every $\kappa,d>0$,
\[
2^{-m}\sum_{\sigma,\tau\in\F^W}e^{-\kappa dN_2N_3/n}\le 10\exp\Big(\frac{200}{\kappa e}\cdot\frac nd\Big).
\]
\end{lemma}

The sum is dominated by colourings that are almost two-coloured. When
$m\le n/2$, colour $0$ is the largest class, and the bound reduces to a
weighted binomial sum whose logarithm is of order
$m\,e^{-\Theta(dm/n)}\le O(n/d)$. This is the analytic counterpart of the
heuristic that a set $W$ of size $m$ contains about $m e^{-dm/n}$ vertices
with no neighbour in $W$, which is largest for $m\approx n/d$.

\begin{proposition}[Kernel moment]\label{prop:moment}
For even $n\ge 128$, every $d\ge1$, every $W\subseteq[n]$ and every
diagonal $\Lambda$,
$\ln\E\,2^{\nul(A[W,W]+\Lambda)}\le\ln 10+C_0\,n/d$ with $C_0=200/(\eta e)$.
\end{proposition}

\begin{proof}
Bound every term of \eqref{eq:phid} by $|\varphi|^d$, apply
Lemma~\ref{lem:fourcolour} and then Lemma~\ref{lem:count} with $\kappa=\eta$.
\end{proof}

It remains to pass to the conditioned law and to check that it has large
min-entropy. Both steps are routine once the exponential moment is
available.

\begin{lemma}\label{lem:simple}
For even $n$ and $2d\le n$, $\Prob(\mathcal S)\ge\exp(-C_1d^2\ln n)$ for an
absolute constant $C_1$.
\end{lemma}

\begin{proposition}[Hard regular distribution]\label{prop:regularprior}
If $d\ge2$, $d^3\ln n\le n$ and $n$ is even and large enough, then
\[
\sup_{W,\Lambda}\E_\pi\nul(A_G[W,W]+\Lambda)=O(n/d)
\qquad\text{and}\qquad
\ln\frac1{\max_g\pi(g)}=\Omega(nd\ln n).
\]
\end{proposition}

\begin{proof}[Proof sketch]
For a nonnegative random variable $X$, conditioning on $\mathcal S$ can
increase $\E\,2^X$ by at most a factor $1/\Prob(\mathcal S)$, and Jensen's
inequality gives $\E[X\mid\mathcal S]\le\log\E\,2^X+\log(1/\Prob(\mathcal S))$.
Proposition~\ref{prop:moment} and Lemma~\ref{lem:simple} bound this by
$O(n/d+d^2\log n)=O(n/d)$. For the min-entropy, all $((n-1)!!)^d$ matching
tuples are equally likely and a $d$-regular graph has at most $(d!)^n$
ordered decompositions into perfect matchings, so
$\pi(g)\le(d!)^n/(((n-1)!!)^d\Prob(\mathcal S))$; Stirling's formula and
$\ln d\le\frac13\ln n$ give $\Omega(nd\ln n)$. Details are in
Appendix~\ref{app:regular}.
\end{proof}

\begin{proof}[Proof of Theorem~\ref{thm:regular}]
The upper bound is Proposition~\ref{prop:upper}. For even $n$, insert
Proposition~\ref{prop:regularprior} into \eqref{eq:recipe} with $V=1$:
$N\cdot O(n/d)\ge\Omega(nd\log n)$. For odd $n$, $d$ must be even; put
the hard distribution on $n'=n-d-1$ vertices and a fixed copy of
$K_{d+1}$ on the remaining ones. The state factorizes, the known factor
carries no information, and the estimates above hold with $n'$ in place
of $n$.
\end{proof}

A random distribution is unavoidable here. For a disjoint union of
copies of $K_{d,d}$ the learner can measure one side of each copy in $X$,
obtaining nullity $n/2$; no bound of the form $O(n/d)$ can hold for every
graph and every setting.

\section{Adaptivity and hidden bipartitions}\label{sec:adaptive}

The obstacle in the nonadaptive setting is that the learner cannot find
large independent sets. An adaptive learner can, provided the graph has a
structure that lets earlier outcomes point to them.

\begin{lemma}[Parallel neighbourhood recovery]\label{lem:rows}
Let $F$ be a set known to be independent in an unknown graph of maximum
degree at most $d$, and let $Q=\sum_{j\le d}\binom{n-1}{j}$. Measuring $X$
on $F$ and $Z$ elsewhere on $b$ fresh copies determines all neighbourhoods
$N(v)$, $v\in F$, by exhaustive search, except with probability at most
$|F|\,Q\,2^{-b}$.
\end{lemma}

\begin{proof}
Since $F$ is independent, every $K_v$ with $v\in F$ is a product of
measured observables, and $s_v=\sum_{u\in N(v)}t_u$ on every copy. The
$Z$-outcome vector $t$ is uniform and independent across copies. A wrong
candidate $\Gamma\ne N(v)$ with $\Gamma\cap F=\emptyset$ satisfies the same
equation only if $\sum_{u\in\Gamma\triangle N(v)}t_u=0$, which has
probability $1/2$ on each copy. A union bound over at most $Q$ candidates
and $|F|$ vertices concludes.
\end{proof}

Compared with Proposition~\ref{prop:upper}, every vertex of $F$ is tested
on every copy, instead of on a fraction $1/d$ of them.

\begin{theorem}[Breadth-first learner]\label{thm:bfs}
Let $G$ be connected and bipartite, with maximum degree at most $d$ and
diameter at most $D$, and with unknown bipartition. An adaptive $X/Z$
learner identifies $G$ with probability at least $1-\delta$ from
$O\big((D+1)[d\ln(en/d)+\ln(n(D+1)/\delta)]\big)$ copies.
\end{theorem}

\begin{proof}
Fix a root and set $b=\lceil\log(nQ(D+1)/\delta)\rceil$. In stage $i$ the
learner knows the set $F_i$ of vertices at distance $i$ from the root; it
spends $b$ fresh copies measuring $X$ on $F_i$ and $Z$ elsewhere, recovers
the neighbourhoods of $F_i$ by Lemma~\ref{lem:rows}, and lets $F_{i+1}$ be
the newly discovered vertices. In a bipartite graph all vertices at the
same distance from the root lie on the same side, so every $F_i$ is
independent and Lemma~\ref{lem:rows} applies, conditionally on all earlier
stages having succeeded. Each edge joins two consecutive layers and is
found from its endpoint in the earlier one, so the last layer is never
queried and at most $D$ stages are used. Each stage fails with
probability at most $\delta/(D+1)$, and $\ln Q=O(d\ln(en/d))$.
\end{proof}

With $D=D_n$ and $\delta=1/3$ this is the adaptive bound of
Theorem~\ref{thm:bipartite}. The factor $D$ is the cost of discovering the
bipartition, as the next statement shows.

\begin{proposition}\label{prop:revealed}
In the parameter range of Theorem~\ref{thm:bipartite}, if the bipartition
of the target is revealed, then $\Theta(d\log n)$ copies are necessary and
sufficient. The upper bound repeats a single nonadaptive $X/Z$ setting;
the lower bound holds even for adaptive Pauli learners.
\end{proposition}

\begin{proof}
Measure $X$ on one side $L$ and $Z$ on the other side, which is
independent, on $O(d\log n)$ copies; Lemma~\ref{lem:rows} recovers all
neighbourhoods of $L$, and every edge has an endpoint in $L$. For the
lower bound, fix the bipartition and use the bijection distribution of
Section~\ref{sec:bipartite} with this bipartition; its largest atom is at
most $(d!)^m/((m!)^d\Prob(\mathcal E))$ with $\Prob(\mathcal E)\ge\frac12e^{-d^2}$,
so its min-entropy is $\Omega(nd\log n)$. Any protocol, adaptive or not,
gains at most $n$ bits per copy, and Lemma~\ref{lem:fano} gives
$N=\Omega(d\log n)$.
\end{proof}

\section{The nonadaptive lower bound on the bipartite family}\label{sec:bipartite}

The distribution of Section~\ref{sec:regular} is supported on graphs with
odd cycles, so it says nothing about $\Bnd$. We use instead the following
distribution. Let $n=2m$. Choose a uniformly random balanced partition
$[n]=L\sqcup R$ and $d$ independent uniform bijections $L\to R$, viewed as
$m\times m$ permutation matrices $P_1,\dots,P_d$, and let $\mathcal S$ be the
event that their edges are pairwise disjoint. On $\mathcal S$ the union is
a simple $d$-regular bipartite graph. A van der Waerden permanent argument
gives $\Prob(\mathcal S)\ge e^{-d^2}$ (Lemma~\ref{lem:bsimple}). Let
$\mathcal E\subseteq\mathcal S$ be the intersection of $\mathcal S$ with the
expansion event of Lemma~\ref{lem:expansion}; its graphs are connected,
have diameter at most $D_n$, and $\Prob(\mathcal E)\ge\frac12e^{-d^2}$. Let
$\pi_B$ be the law of the graph conditioned on $\mathcal E$, forgetting the
bipartition and the matchings.

\paragraph{Where the nullity comes from.}
For a fixed $W$ let $w_L=|W\cap L|$, $w_R=|W\cap R|$, $k=\min(w_L,w_R)$ and
$\ell=\max(w_L,w_R)$. Since all edges go between $L$ and $R$,
\begin{equation}\label{eq:blocknul}
A_G[W,W]=\begin{pmatrix}0&B\\ B^{\mathsf T}&0\end{pmatrix},\qquad
\nul A_G[W,W]=|w_L-w_R|+2\,(k-\rank B),
\end{equation}
where $B$ is the $\ell\times k$ biadjacency block (after transposing if
necessary). The first term is what a learner gains by knowing the sides:
with $W=L$ it equals $n/2$. Because the partition is uniform and hidden,
$\E|w_L-w_R|\le\sqrt{n}$ for every fixed $W$. The second term is a rank
deficiency of a rectangular block of a sum of random permutation matrices.

With $Y$ measurements a diagonal $\Lambda$ is added and \eqref{eq:blocknul}
no longer applies. We deal with this by comparing with an unshifted sum.

\begin{lemma}[Fixed perturbations]\label{lem:shift}
Let $X_1,X_2,\dots$ be independent identically distributed $w\times w$
matrices over $\F$, let $d\ge2$ and $d'=2\lfloor d/2\rfloor$. For every fixed
$w\times w$ matrix $T$,
\[
\E\,2^{\nul(X_1+\dots+X_d+T)}\le\E\,2^{\nul(X_1+\dots+X_{d'})}.
\]
\end{lemma}

\begin{proof}
Identity \eqref{eq:kernelcount} holds for any square matrix (the kernel
of the transpose has the same size). With
$\psi(x,y)=\E(-1)^{x^{\mathsf T}X_1y}$, which is real and of absolute value at
most one,
\[
\E\,2^{\nul(X_1+\dots+X_d+T)}=2^{-w}\sum_{x,y}(-1)^{x^{\mathsf T}Ty}\psi(x,y)^d
\le 2^{-w}\sum_{x,y}|\psi(x,y)|^{d'}=2^{-w}\sum_{x,y}\psi(x,y)^{d'} ,
\]
using $d'\le d$ and that $d'$ is even. The last sum equals the right-hand
side.
\end{proof}

The evenness matters: if every $X_j$ equals $\left(\begin{smallmatrix}0&1\\1&0\end{smallmatrix}\right)$,
an odd sum is invertible, while adding the identity produces nullity
one. Applied to the bipartite distribution before conditioning, with a
fixed partition, Lemma~\ref{lem:shift} and \eqref{eq:blocknul} give
\[
\E\,2^{\nul(A[W,W]+\Lambda)}\le 2^{|w_L-w_R|}\,\E\,4^{k-\rank B_{d'}},
\]
where $B_{d'}$ is the corresponding block of a sum of $d'$ permutation
matrices. We are thus led to a second moment of the rectangular kernel.

\begin{lemma}[Second rectangular moment]\label{lem:second}
Let $B$ be a fixed $\ell\times k$ submatrix of the sum over $\F$ of $d\ge1$
independent uniform $m\times m$ permutation matrices, $k\le\ell\le m$.
With $c_0=(\ln2-\ln(1+e^{-4}))/4$,
\[
\ln\E\,4^{k-\rank B}\le 4d\ln(m+1)+4\ln(k+1)+\frac{2m}{e\,c_0\,d}.
\]
\end{lemma}

Here $4^{k-\rank B}=|\ker B|^2$ counts pairs $(y,z)$ of kernel vectors, and
we encode each column $j$ by the symbol $(y_j,z_j)\in\F^2$, an echo of the
four colours of Section~\ref{sec:regular}. A permutation matrix acts on
this word of symbols by a uniform shuffle. Shuffles are awkward because
their entries are dependent, but a shuffle is just a sequence of
independent samples conditioned on its exact counts, and the exact counts
are a mode of the multinomial law, which has probability at least
$(m+1)^{-4}$. Paying this factor once per permutation, the rows become
independent, and the probability that a row sums to zero is a convolution
on the group $\F^2$, which Fourier analysis controls. Balancing the number
of words of a given type against this probability gives the bound. The
proof is in Appendix~\ref{app:bipartite}.

\begin{lemma}[Hard bipartite distribution]\label{lem:bprior}
If $d\ge3$, $d^3\ln n\le n$ and $n$ is even and large enough, then
\[
\ln\frac1{\max_g\pi_B(g)}=\Omega(nd\ln n)
\qquad\text{and}\qquad
\sup_{W,\Lambda}\E_{\pi_B}\nul(A_G[W,W]+\Lambda)=O(n/d).
\]
\end{lemma}

The order of operations in the proof matters. We fix the partition,
compare with the unshifted $d'$-sum, apply Lemma~\ref{lem:second},
condition on $\mathcal E$ at a cost $\ln(2e^{d^2})$, and take logarithms
with Jensen's inequality. Only then do we average over the hidden
partition, which turns $|w_L-w_R|$ into $O(\sqrt n)$; averaging
$2^{|w_L-w_R|}$ instead would be far too costly.

\begin{lemma}[Expansion and diameter]\label{lem:expansion}
Let $\lceil\ln n\rceil\le d$, $d^3\ln n\le n$, $\alpha=\lfloor\ln n/32\rfloor$,
$r_0=\lfloor m/(4\alpha)\rfloor$ and $q_0=\lfloor m/5\rfloor$. Conditionally on
$\mathcal S$, with probability $1-o(1)$, every set $U$ on one side with
$1\le|U|\le r_0$ has at least $\alpha|U|$ neighbours, and every two sets of
size $q_0$ on opposite sides are joined by an edge. Every simple graph
with these two properties is connected and has diameter at most
$16\ln n/\ln\ln n$ for large $n$.
\end{lemma}

The two properties are checked for independent bijections by union
bounds and then transferred to the conditioned law by dividing by
$\Prob(\mathcal S)\ge e^{-d^2}$, which is affordable because the failure
probabilities are of order $e^{-\Omega(d^2\ln n)}$ and $e^{-\Omega(dm)}$.
Balls around any vertex grow by a factor $\alpha\approx\ln n/32$ per step until
they reach size $m/5$, and the second property joins any two such balls.
This is where $d\ge\ln n$ is needed.

\begin{proof}[Proof of Theorem~\ref{thm:bipartite}]
By Lemma~\ref{lem:expansion}, $\pi_B$ is supported on $\Bnd$, and
\eqref{eq:recipe} with Lemma~\ref{lem:bprior} gives the nonadaptive lower
bound. The nonadaptive upper bound is Proposition~\ref{prop:upper}, since
$\Bnd\subseteq\Rnd$, and the adaptive bound is Theorem~\ref{thm:bfs} with
$D=D_n$.
\end{proof}

\section{Approximate recovery}\label{sec:approx}

\begin{proof}[Proof of Theorem~\ref{thm:approx}]
Let $M=\binom n2$. Around any simple graph, the number of graphs within
edge distance $R\ge1$ is $\sum_{j\le R}\binom Mj\le(eM/R)^R$. Both hard
distributions satisfy $\ln(1/\pi_*)\ge c\,nd\ln n$ for an absolute $c>0$
(for odd $n$ in the regular case, through the clique construction). Take
$R=\lfloor\varepsilon nd\rfloor$ with $\varepsilon\le\varepsilon_0$ and $2\varepsilon_0<c/2$.
Since $x\mapsto x\ln(eM/x)$ is increasing on $(0,M]$, the ball has
logarithmic size at most $\varepsilon_0 nd\ln(eM/(\varepsilon_0nd))\le 2\varepsilon_0nd\ln n$ for
large $n$, so $\ln(1/(V\pi_*))\ge(c/2)nd\ln n$ in Lemma~\ref{lem:fano}.
The information upper bound $O(Nn/d)$ is unchanged, which gives
$N=\Omega(d^2\log n)$; exact recovery meets the tolerance, so the upper
bounds still apply.
\end{proof}

In words, allowing a small constant fraction of wrong edges only removes
the logarithm of a ball from the min-entropy, which is a small fraction of
it. We stress that the loss here is on the edge set; state fidelity,
prediction error under random test measurements, and noisy input copies
are different questions.

\section{Discussion}\label{sec:discussion}

The nonadaptive results identify a clean mechanism: a local Pauli
measurement extracts information only through stabilizers it happens to
align with, and without knowledge of the graph only about $n/d$ of them
can be aligned on a sparse random graph. Adaptivity helps when earlier
outcomes reveal large independent sets, as in the bipartite case. Two
small additions in the appendix complement this picture. On the three
labelled four-cycles, two adaptive copies identify the graph with
probability $41/48$, while the best two nonadaptive copies reach $5/6$
(Appendix~\ref{app:finite}). And for adaptive separable measurements, the
information per copy is at most $n$ minus the geometric entanglement of the
state (Appendix~\ref{app:separable}), which suggests a route towards
adaptive lower bounds on $\Rnd$.

Several questions remain open. The optimal adaptive rate is unknown both on
$\Bnd$ and on $\Rnd$; on $\Rnd$ there are no independent breadth-first
layers to exploit, and we do not know whether adaptivity helps at all.
Adaptivity within a single copy, with classical feed-forward as in
measurement-based computation, is a different resource from the
between-copy adaptivity studied here. Our lower bounds cover Pauli
measurements; arbitrary single-qubit bases would require a replacement
for the affine outcome law. The degree range $d^3\ln n\le n$ comes from the
cost of conditioning on disjoint matchings and could perhaps be relaxed.
Finally, all our learners use exhaustive search over candidate
neighbourhoods and take time $n^{O(d)}$; whether $\mathrm{poly}(d)\log n$
copies can be achieved in polynomial time is an interesting
computational question.

\bibliographystyle{plainnat}
\bibliography{refs}

@inproceedings{ot2022,
  author    = {Ouyang, Yingkai and Tomamichel, Marco},
  title     = {Learning quantum graph states with product measurements},
  booktitle = {IEEE International Symposium on Information Theory (ISIT)},
  pages     = {2963--2968},
  year      = {2022},
  note      = {arXiv:2205.06432}
}

@inproceedings{ms2022,
  author    = {Montanaro, Ashley and Shao, Changpeng},
  title     = {Quantum algorithms for learning a hidden graph},
  booktitle = {17th Conference on the Theory of Quantum Computation, Communication and Cryptography (TQC)},
  series    = {LIPIcs},
  volume    = {232},
  pages     = {1:1--1:22},
  year      = {2022}
}

@article{hein2004,
  author  = {Hein, Marc and Eisert, Jens and Briegel, Hans J.},
  title   = {Multiparty entanglement in graph states},
  journal = {Physical Review A},
  volume  = {69},
  pages   = {062311},
  year    = {2004}
}

@inproceedings{abdy2023,
  author    = {Arunachalam, Srinivasan and Bravyi, Sergey and Dutt, Arkopal and Yoder, Theodore J.},
  title     = {Optimal algorithms for learning quantum phase states},
  booktitle = {18th Conference on the Theory of Quantum Computation, Communication and Cryptography (TQC)},
  year      = {2023},
  note      = {arXiv:2208.07851}
}

@misc{as2026,
  author = {Arunachalam, Srinivasan and Schatzki, Louis},
  title  = {Optimal stabilizer testing and learning with limited quantum memory},
  year   = {2026},
  note   = {arXiv:2607.02444}
}

@misc{begms2026,
  author = {Bittel, Lennart and Eisert, Jens and Gong, Weiyuan and Mele, Antonio A. and Schatzki, Louis},
  title  = {Adaptivity is all you need: Optimal stabilizer learning using just single-copy measurements},
  year   = {2026},
  note   = {arXiv:2610.02031}
}

@misc{ln2022,
  author = {Lowe, Angus and Nayak, Ashwin},
  title  = {Lower bounds for learning quantum states with single-copy measurements},
  year   = {2022},
  note   = {arXiv:2207.14438}
}

@misc{lto2026,
  author = {Leitch, Heather and Tirukkovalluri, Sri S. and Ouyang, Yingkai},
  title  = {Learning unknown stabilizer codes using product measurements},
  year   = {2026},
  note   = {arXiv:2609.04997}
}

@article{rocchetto2018,
  author  = {Rocchetto, Andrea},
  title   = {Stabiliser states are efficiently {PAC}-learnable},
  journal = {Quantum Information and Computation},
  volume  = {18},
  pages   = {541--552},
  year    = {2018}
}

@inproceedings{adgo2026,
  author    = {Arunachalam, Srinivasan and Dutt, Arkopal and Gheorghiu, Alexandru and de Oliveira, Michael},
  title     = {Learning depth-3 circuits via quantum agnostic boosting},
  booktitle = {Conference on Learning Theory (COLT)},
  year      = {2026},
  note      = {arXiv:2509.14461}
}

@inproceedings{ab2019,
  author    = {Abasi, Hasan and Bshouty, Nader H.},
  title     = {On learning graphs with edge-detecting queries},
  booktitle = {Algorithmic Learning Theory (ALT)},
  year      = {2019},
  note      = {arXiv:1803.10639}
}

@article{meszaros2020,
  author  = {M{\'e}sz{\'a}ros, Andr{\'a}s},
  title   = {The distribution of sandpile groups of random regular graphs},
  journal = {Transactions of the American Mathematical Society},
  volume  = {373},
  pages   = {6529--6594},
  year    = {2020}
}

@misc{portier2026,
  author = {Portier, Julien},
  title  = {Sprinkling for random regular graphs and {Kim--Vu}'s sandwich conjecture},
  year   = {2026},
  note   = {arXiv:2610.28395}
}

@article{ghh2025,
  author  = {Ghosh, Soumik and Hangleiter, Dominik and Helsen, Jonas},
  title   = {Random regular graph states are complex at almost any depth},
  journal = {PRX Quantum},
  year    = {2025},
  note    = {arXiv:2412.07058}
}

@article{egorychev1981,
  author  = {Egorychev, G. P.},
  title   = {The solution of van der {W}aerden's problem for permanents},
  journal = {Advances in Mathematics},
  volume  = {42},
  pages   = {299--305},
  year    = {1981}
}

@article{falikman1981,
  author  = {Falikman, D. I.},
  title   = {Proof of the van der {W}aerden conjecture regarding the permanent of a doubly stochastic matrix},
  journal = {Mathematical Notes of the Academy of Sciences of the USSR},
  volume  = {29},
  pages   = {475--479},
  year    = {1981}
}

\appendix

\section{Proofs for Section~\ref{sec:model}}\label{app:model}

\begin{proof}[Proof of Lemma~\ref{lem:outcome}]
For $a\in\F^n$ the product $\prod_{v:a_v=1}K_v$ equals $(-1)^{q_G(a)}X^aZ^{Aa}$,
where $X^a$ is written to the left of $Z^{Aa}$; this follows by commuting
the factors, or directly from the action on basis states. Take $a$
supported on $W$. On a qubit $i\in W$ this operator acts as $X$ if
$a_i=1,(Aa)_i=0$; as $XZ=-iY$ if $a_i=(Aa)_i=1$; as $Z$ if $a_i=0,(Aa)_i=1$;
and trivially otherwise. On $U$ it acts by $Z$ or trivially. Hence it is,
up to a phase, a product of measured observables exactly when
$(Aa)_i=\Lambda_{ii}a_i$ for all $i\in W$, that is, when $(C+\Lambda)a=0$.

For such $a$, $0=a^{\mathsf T}(C+\Lambda)a=a^{\mathsf T}\Lambda a=\wt(\Lambda a)\bmod 2$,
because $a^{\mathsf T}Ca=0$ for a symmetric matrix with zero diagonal. So
the phase $(-i)^{\wt(\Lambda a)}$ equals $(-1)^{\wt(\Lambda a)/2}$, and since
$|G\rangle$ is a $+1$ eigenvector of the stabilizer, the product of the
corresponding outcomes is $(-1)^{\omega(a)}$. Its $W$-part is $a$ and its
$U$-part is $B^{\mathsf T}a$, which is \eqref{eq:affine}.

Linearity of $\omega$ on $K$ follows from $q(a+b)=q(a)+q(b)+a^{\mathsf T}Cb$
and from $\wt(\Lambda(a+b))=\wt(\Lambda a)+\wt(\Lambda b)-2|\supp\Lambda
a\cap\supp\Lambda b|$, which give
$\omega(a+b)=\omega(a)+\omega(b)+a^{\mathsf T}(C+\Lambda)b$.

Any other parity of the outcomes is the eigenvalue of a Pauli operator
$P$ that is not $\pm$ a stabilizer. Such a $P$ anticommutes with some
generator $K_v$, so $\langle P\rangle=\langle PK_v\rangle=-\langle
K_vP\rangle=-\langle P\rangle=0$. By Fourier inversion on $\F^n$,
\[
\Prob(s,t)=2^{-n}\sum_{a\in K}(-1)^{\omega(a)+a\cdot s+a^{\mathsf T}Bt}.
\]
The exponent is linear in $a\in K$, so the sum is $|K|$ when
\eqref{eq:affine} holds and $0$ otherwise. The equations are independent
because their coefficients on $s$ are the basis vectors of $K$, which
gives the dimension count.
\end{proof}

For $X/Z$ settings there is also a direct computation. After the
$Z$ outcomes $t$ are fixed, the $W$ qubits are in the state
$Z^{Bt}|G[W]\rangle$, up to a global sign. The amplitude of the $X$ outcome
$s$ is proportional to $F(\ell)=\sum_x(-1)^{q_W(x)+\ell\cdot x}$ with
$\ell=s+Bt$, and substituting $x'=x+a$ in $F(\ell)^2$ together with
$q_W(x)+q_W(x+a)=q_W(a)+x^{\mathsf T}Ca$ gives
$F(\ell)^2=2^m\sum_{a\in\ker C}(-1)^{q_W(a)+\ell\cdot a}$.

\begin{proof}[Proof of Proposition~\ref{prop:upper}]
Put $q=1/(\Delta+1)$. For each vertex $v$ keep every candidate
neighbourhood $\Gamma\subseteq[n]\setminus\{v\}$ with $|\Gamma|\le\Delta$, and
discard $\Gamma$ whenever, on some copy, $v$ is measured in $X$, all of
$\Gamma$ in $Z$, and $s_v+\sum_{u\in\Gamma}t_u=1$. The true neighbourhood is
never discarded, because this parity is the value of $K_v$. For
$\Gamma\ne N(v)$ the operator $X_vZ_\Gamma$ is not $\pm$ a stabilizer (the only
stabilizer with $X$-part $\{v\}$ is $K_v$), so the parity is a fair coin
whatever the other qubits are measured in. A wrong $\Gamma$ is therefore
discarded on each copy with probability at least
$\frac12q(1-q)^\Delta\ge1/(4e\Delta)$. There are at most
$\sum_{j\le\Delta}\binom{n-1}j\le(en/\Delta)^\Delta$ candidates per vertex
(use $t^{-\Delta}(1+t)^{n-1}$ with $t=\Delta/n$), and a union bound over the
$n(en/\Delta)^\Delta$ candidates finishes the proof.
\end{proof}

\section{Proofs for Section~\ref{sec:regular}}\label{app:regular}

\begin{proof}[Proof of Lemma~\ref{lem:fourcolour}]
\emph{Symmetry.} Over $\F$ the quantity $c_i\wedge c_j=\sigma_i\tau_j+\sigma_j\tau_i$ is the
determinant of the matrix with columns $c_i,c_j$, so it is invariant
under every invertible linear map of $\F^2$. Translating all colours by
$r$ adds $r\wedge\sum_ic_i$ to the exponent, the same for every matching.
Hence $|\varphi|$ is invariant under all $24$ affine permutations of $\F^2$,
that is, under all permutations of the four colours. We may assume that
the largest class has colour $0$, so $N_1\ge n/4$, and that the next two
have colours $A\ne B$. If $N_3=0$ there is nothing to prove. Put
$\mu=N_2N_3/n$.

\emph{Blocks.} Generate $M$ by ordering the vertices uniformly at random,
cutting the order into $q=\lfloor n/4\rfloor$ blocks of four (and one pair
if $n\equiv2\pmod 4$), and choosing one of the three internal matchings of
each block uniformly and independently. The result is a uniform perfect
matching, since the construction does not depend on vertex names. Call a
block good if its colours are $0,0,A,B$. Its three internal matchings
have signs $-1,+1,+1$, so its average sign is $1/3$; any other block has
average sign of absolute value at most one. Conditioning on the order, the
sign is a product over blocks of independent factors, so
$|\varphi|\le\E\,3^{-L}$, where $L$ is the number of good blocks.

\emph{Large $\mu$.} Expose the first $k=\lfloor n/64\rfloor$ blocks one at a
time and stop as soon as one of the classes $0,A,B$ has lost more than
half of its vertices. Before stopping, the remaining class sizes are at
least $n/8$ (so the remaining zero class $z'$ satisfies $z'-1\ge n/16$),
$N_2/2$ and $N_3/2$, so the next block is good with conditional probability
at least $12z'(z'-1)\frac{N_2}2\frac{N_3}2/n^4\ge p:=3N_2N_3/(128n^2)$. Comparing
with independent Bernoulli$(p)$ trials,
$\E[3^{-L};\text{no stop}]\le(1-2p/3)^k\le e^{-\mu/8192}$. A class of size
$r$ can only lose half of its vertices if at least $\lceil r/2\rceil$ of
them fall among the first $4k\le n/16$ positions, which has probability
at most $\binom r{\lceil r/2\rceil}16^{-\lceil r/2\rceil}\le2^{-r}$. As $\mu\le N_3$,
we get $\E\,3^{-L}\le e^{-\mu/8192}+3\cdot2^{-N_3}\le4e^{-\mu/8192}$.

\emph{Small $\mu$.} Over all $q$ blocks, sampling without replacement gives
\[
\E L=\frac{12\,q\,(N_1)_2N_2N_3}{(n)_4},\qquad
\E L(L-1)=\frac{144\,q(q-1)\,(N_1)_4(N_2)_2(N_3)_2}{(n)_8},
\]
where $(x)_j$ is the falling factorial. For $n\ge128$ these yield $3\mu/64\le\E L\le6\mu$ and
$\E L^2\le6\mu+18\mu^2$, hence $\Prob(L>0)\ge(\E L)^2/\E L^2\ge\mu/(8192(1+\mu))$.
Since $3^{-L}\le1-\frac23\mathbf 1\{L>0\}$, we get
$1-\E\,3^{-L}\ge\mu/(12288(1+\mu))$, which is at least $\eta\mu$ when
$\mu\le32768$, because $12288\cdot32769<2^{29}$. Then $\E\,3^{-L}\le1-\eta\mu\le
e^{-\eta\mu}$. For $\mu\ge32768$ the previous paragraph gives
$\E\,3^{-L}\le4e^{-\mu/8192}\le e^{-\mu/16384}\le e^{-\eta\mu}$.
\end{proof}

The restriction to large $n$ is real: for $n=4$ with four distinct colours
every matching has sign $-1$.

\begin{proof}[Proof of Lemma~\ref{lem:count}]
Write $T$ for the left-hand side and $h$ for the binary entropy in nats.

\emph{Case $m\le n/2$.} Colour $0$ is the largest class, since all
$n-m\ge n/2$ vertices outside $W$ carry it. Fix a colouring with $k$
nonzero vertices in $W$, let $a'$ be the size of the largest nonzero class
and $r=k-a'$. Then $N_2=a'\ge k/3$ and $N_3\ge r/2$, so $N_2N_3\ge kr/6$.
Choosing the positions of the nonzero vertices, designating one of the
three nonzero colours as dominant, and letting each of the $k$ positions
either take it or one of the two others, we obtain
\[
T\le3\cdot2^{-m}\sum_{k=0}^m\binom mk\big(1+2e^{-\kappa dk/(6n)}\big)^k .
\]
Terms with $k\le m/8$ contribute at most $3e^{-\gamma m}\le3$, with
$\gamma=\ln2-h(1/8)-\frac18\ln3>0$, by the bound
$\sum_{k\le m/8}\binom mk3^k\le x^{-m/8}(1+3x)^m$ at $x=1/21$. For $k>m/8$
replace the exponential by $u=e^{-\kappa dm/(48n)}$ and complete the binomial
sum: $2^{-m}\sum_k\binom mk(1+2u)^k=(1+u)^m\le e^{mu}$. So
$T\le6\exp(m e^{-\kappa dm/(48n)})\le6\exp(48n/(\kappa ed))$, using $xe^{-x}\le1/e$.

\emph{Case $m>n/2$.} If $N_2<n/100$, all colours but the dominant one
occupy fewer than $3m/50$ positions of $W$; choosing the dominant colour
in four ways, these colourings contribute at most
$4\exp\{m[h(3/50)+\frac3{50}\ln3-\ln2]\}\le4$. Otherwise $N_2\ge n/100$; with
$r=N_3+N_4$ we have $N_3\ge r/2$ and the penalty is at most $e^{-\kappa dr/200}$.
Choosing the pair of dominant colours in six ways and letting $\ell\le r$
positions of $W$ lie outside the pair, these colourings contribute at most
$6\cdot2^{-m}\sum_\ell\binom m\ell2^me^{-\kappa d\ell/200}=6(1+e^{-\kappa d/200})^m\le
6\exp(200n/(\kappa ed))$. Together, $T\le10\exp(200n/(\kappa ed))$.
\end{proof}

\begin{lemma}\label{lem:avoid}
Let $H$ be a graph of maximum degree at most $r\ge1$ on an even number
$n\ge2r+2$ of vertices. A uniform perfect matching avoids all edges of $H$
with probability at least $\exp(-C r\ln n)$ for an absolute constant $C$.
\end{lemma}

\begin{proof}
Generate the matching by repeatedly pairing the smallest unmatched vertex
with a uniformly random unmatched partner; this produces the uniform
distribution, and the remaining pairs are always uniform given the
earlier ones. While $k>K=2r+2$ vertices remain, the next pair avoids $H$
with probability at least $1-r/(k-1)\ge e^{-2r/(k-1)}$, and the product of
these factors is at least $e^{-2r(\ln n+1)}$. On the last $K$ vertices the
allowed pairs form a graph of minimum degree at least $K-1-r=K/2$, which
has a perfect matching: in a maximum matching missing two vertices $u,v$,
each matched pair $\{x,y\}$ contributes at most two to $\deg u+\deg v$,
since $u\sim x$ and $v\sim y$ would give an augmenting path, so $\deg
u+\deg v\le K-2$, a contradiction. The random completion hits an allowed
one with probability at least $1/(K-1)!!\ge e^{-C'r\ln n}$.
\end{proof}

\begin{proof}[Proof of Lemma~\ref{lem:simple}]
Given that $M_1,\dots,M_{j-1}$ are disjoint, their union has maximum degree
$j-1$ and $M_j$ is independent of it, so by Lemma~\ref{lem:avoid} it avoids
the union with probability at least $e^{-C(j-1)\ln n}$. Multiply over
$j\le d$.
\end{proof}

\begin{proof}[Proof of Proposition~\ref{prop:regularprior}]
Fix $W$ and $\Lambda$ and write $X=\nul(A[W,W]+\Lambda)$. On $\mathcal S$ the
matrix $A$ is the adjacency matrix of the union graph, and
$\E[2^X\mid\mathcal S]\le\E\,2^X/\Prob(\mathcal S)$ because $2^X\ge0$. By Jensen's
inequality and Proposition~\ref{prop:moment} and Lemma~\ref{lem:simple},
\[
\E[X\mid\mathcal S]\le\log\E\,2^X+\log\frac1{\Prob(\mathcal S)}=O\Big(\frac nd+d^2\log n\Big)=O\Big(\frac nd\Big),
\]
since $d^3\ln n\le n$. For the min-entropy, the $((n-1)!!)^d$ ordered
matching tuples are equally likely. An ordered decomposition of a
$d$-regular graph into perfect matchings is determined by a bijection
between the $d$ edges and the $d$ colours at every vertex, so there are
at most $(d!)^n$ of them. Hence
$\pi(g)\le(d!)^n/(((n-1)!!)^d\Prob(\mathcal S))$ and, by Stirling,
\[
\ln\frac1{\pi(g)}\ge\frac{nd}2\ln n-nd\ln d-O(nd+d^2\ln n)\ge\Big(\frac16-o(1)\Big)nd\ln n,
\]
because $\ln d\le\frac13\ln n$ in our range.
\end{proof}

\section{Proofs for Section~\ref{sec:bipartite}}\label{app:bipartite}

\begin{lemma}\label{lem:bsimple}
For $d\le m/2$, $\Prob(\mathcal S)\ge e^{-d^2}$.
\end{lemma}

\begin{proof}
After $r$ disjoint bijections have been chosen, the allowed pairs form an
$(m-r)$-regular bipartite graph. Its biadjacency matrix divided by $m-r$
is doubly stochastic, so by the van der Waerden bound
\citep{egorychev1981,falikman1981} it has at least $(m-r)^m m!/m^m$
perfect matchings, and a uniform bijection is allowed with probability at
least $(1-r/m)^m$. Hence $\ln\Prob(\mathcal S)\ge m\sum_{r<d}\ln(1-r/m)\ge-2\sum_{r<d}r\ge-d^2$.
\end{proof}

\begin{lemma}[Types]\label{lem:types}
Let $p$ be a distribution on at most four symbols with integer counts
$r_i=mp_i$. Then $m$ independent samples from $p$ have exactly the counts
$(r_i)$ with probability at least $(m+1)^{-4}$, and conditionally on these
counts the sample word is uniform among the arrangements of a word with
counts $(r_i)$.
\end{lemma}

\begin{proof}
Every word with given counts has the same probability, which proves the
second claim. For the first, $(r_i)$ is a mode of the multinomial law:
if a count vector $b$ has $b_i>r_i$ and $b_j<r_j$, moving one unit from
$i$ to $j$ multiplies its probability by $\frac{b_i}{b_j+1}\cdot\frac{r_j}{r_i}>1$.
There are at most $(m+1)^4$ count vectors.
\end{proof}

\begin{lemma}\label{lem:scalar}
Let $h(p)=-p\ln p-(1-p)\ln(1-p)$ and $f_\gamma(p)=h(p)-\ln2+\ln(1+e^{-\gamma p})$.
For $\gamma\ge0$ and $0\le p\le1/2$, $f_\gamma(p)\le e^{-c_0\gamma}$.
\end{lemma}

\begin{proof}
If $p\ge1/4$, then $f_\gamma(p)\le\ln(1+e^{-\gamma/4})\le e^{-\gamma/4}\le e^{-c_0\gamma}$.
If $p\le1/4$ and $\gamma p\ge4$, then $f_\gamma(p)\le h(1/4)-\ln2+\ln(1+e^{-4})<0$.
Otherwise the convex function $t\mapsto\ln((1+e^{-t})/2)$ lies below its
chord on $[0,4]$, which is $-c_0t$, so
$f_\gamma(p)\le h(p)-c_0\gamma p\le p(\ln(e/p)-c_0\gamma)\le e^{-c_0\gamma}$, the last
step by maximizing over $p$.
\end{proof}

\begin{proof}[Proof of Lemma~\ref{lem:second}]
For $k=0$ there is nothing to prove. Otherwise
$4^{k-\rank B}=|\ker B|^2=\sum_{y,z\in\F^k}\mathbf 1\{By=Bz=0\}$. Fix
$(y,z)$, encode column $j$ by the symbol $(y_j,z_j)\in\F^2$, let $q$ be the
empirical distribution of these $k$ symbols, and extend the word by zero
symbols to all $m$ columns; its empirical distribution is
$p=(1-\theta)\delta_0+\theta q$ with $\theta=k/m$. Row $i$ of $(By,Bz)$ is the sum
over the $d$ permutations of the symbol that each sends to row $i$. Each
permutation shuffles the word uniformly, and by Lemma~\ref{lem:types} a
uniform shuffle is a sequence of $m$ independent samples from $p$
conditioned on an event of probability at least $(m+1)^{-4}$. Removing the
$d$ conditionings multiplies probabilities by at most $(m+1)^{4d}$, and
without them the rows are independent, so
\[
\Prob(By=Bz=0)\le(m+1)^{4d}\beta^\ell,\qquad\beta=p^{*d}(0),
\]
the probability that a sum of $d$ independent samples from $p$ vanishes in
$\F^2$.

For a nontrivial character $\chi$ of $\F^2$ let $u_\chi=q(\chi=-1)$ and
$\rho_\chi=\min(u_\chi,1-u_\chi)$. The Fourier coefficient of $p$ at $\chi$ is
$1-2\theta u_\chi$, and $|1-2\theta u|\le1-2\theta\min(u,1-u)\le e^{-2\theta\min(u,1-u)}$. By
Fourier inversion, with $\gamma=d\theta$,
\[
\beta\le\frac14\Big(1+\sum_{\chi\ne1}e^{-2\gamma\rho_\chi}\Big).
\]
Both $H(q)$ and the multiset $\{\rho_\chi\}$ are invariant under the affine
permutations of $\F^2$, so we may compute them after relabelling the
masses $q_1\ge q_2\ge q_3\ge q_4$ as $00,01,10,11$. Then the minority masses
are $v=q_3+q_4$, $w=q_2+q_4$ and $z=\min(q_1+q_4,q_2+q_3)$, with
$0\le v\le w\le1/2$ and $z\ge w$. Subadditivity of entropy over the two bits
gives $H(q)\le h(v)+h(w)$. With $a=e^{-\gamma v}\ge b=e^{-\gamma w}$ we have
$1+a^2+2b^2\le(1+a)(1+b)$, because the difference is
$a(1-a)+b(1+a-2b)\ge0$. Therefore, by Lemma~\ref{lem:scalar},
\[
H(q)+\ln\beta\le f_\gamma(v)+f_\gamma(w)\le2e^{-c_0\gamma}.
\]
There are at most $e^{kH(q)}$ words of type $q$, and $\ell\ge k$, so the
pairs $(y,z)$ of type $q$ contribute at most
$(m+1)^{4d}\exp(2k e^{-c_0dk/m})\le(m+1)^{4d}\exp(2m/(ec_0d))$. Summing
over at most $(k+1)^4$ types proves the lemma.
\end{proof}

\begin{proof}[Proof of Lemma~\ref{lem:bprior}]
\emph{Min-entropy.} Before conditioning there are $\binom{2m}m(m!)^d$
equally likely partition and bijection tuples. A connected bipartite
graph has two ordered bipartitions and, for each, at most $(d!)^m$
ordered decompositions into bijections. Hence
$\ln(1/\pi_B(g))\ge d\ln(m!)-m\ln(d!)-d^2-O(n)=\Omega(nd\ln n)$, using
$\ln d\le\frac13\ln n$.

\emph{Nullity.} The event $\mathcal E$ is invariant under relabelling the
vertices, so $\Prob(\mathcal E\mid L,R)$ does not depend on the partition and
the partition remains uniform under $\pi_B$. Fix the partition, $W$ and
$\Lambda$, and let $d'=2\lfloor d/2\rfloor\ge d/2$. Lemma~\ref{lem:shift}, applied
to the independent restrictions of the bijections to $W\times W$, and
\eqref{eq:blocknul} for the unshifted $d'$-sum give
$\E\,2^{\nul(A[W,W]+\Lambda)}\le2^{|w_L-w_R|}\E\,4^{k-\rank B_{d'}}$.
Conditioning on $\mathcal E$ multiplies the left-hand side by at most
$2e^{d^2}$. Jensen's inequality and Lemma~\ref{lem:second} give
\[
\E[\nul(A_G[W,W]+\Lambda)\mid\mathcal E,L,R]\le|w_L-w_R|+O\Big(\frac md+d\ln m+d^2\Big).
\]
Only now average over the uniform partition. The hypergeometric
variance of $w_L$ gives $\E|w_L-w_R|\le\sqrt n$, and in our range $\sqrt n$,
$d\ln m$ and $d^2$ are all $O(n/d)$.
\end{proof}

\begin{proof}[Proof of Lemma~\ref{lem:expansion}]
\emph{Small sets.} In a simple $d$-regular graph every nonempty set has at
least $d\ge\alpha|U|$ neighbours when $|U|\le d/\alpha$. For
$d/\alpha<k\le r_0$, failure means that some $k$-set is mapped by all $d$
bijections into some $\alpha k$-set. Before conditioning, the probability is
at most $2\binom mk\binom m{\alpha k}(\alpha k/m)^{dk}\le2\exp(-\frac d2k\ln\frac m{\alpha k})$,
using $\alpha\le d/32$ and $m/(\alpha k)\ge4$. The exponent is smallest at the
smallest $k$, where $k\ge d/\alpha$ and $\alpha k\le2d$, so summing over $k$
gives at most $2m\exp(-\frac{d^2}{2\alpha}\ln\frac m{2d})\le2me^{-8d^2}$, since
$\ln(m/(2d))\ge\frac12\ln n$ and $\alpha\le\ln n/32$. Dividing by
$\Prob(\mathcal S)\ge e^{-d^2}$ leaves $2me^{-7d^2}=o(1)$ because $d\ge\ln n$.

\emph{No holes.} For fixed sets $X\subseteq L$, $Y\subseteq R$ of size $q_0$, a
uniform bijection maps $X$ into $R\setminus Y$ with probability at most
$(1-q_0/m)^{q_0}\le e^{-q_0^2/m}$. A union bound over at most $4^m$ pairs and
division by $\Prob(\mathcal S)$ give at most $\exp(d^2+2m\ln2-dq_0^2/m)=o(1)$.

\emph{Diameter.} For a vertex $v$ let $S_j$ be the set of endpoints of
walks of length $j$ from $v$; these sets alternate between the sides and
$S_{j+1}=N(S_j)$. While $|S_j|\le r_0$ they grow by a factor at least $\alpha$;
once $|S_j|\ge r_0$, an $r_0$-subset already has at least $\alpha r_0\ge q_0$
neighbours, and from then on all sets have size at least $q_0$. This
happens after at most $t=\lceil\ln r_0/\ln\alpha\rceil+1$ steps. For two
vertices, choose walk lengths in $\{t,t+1\}$ so that their sets lie on
opposite sides; the second property joins them by an edge. So the
distance is at most $2t+3\le16\ln n/\ln\ln n$ for large $n$, as
$\ln\alpha\sim\ln\ln n$.
\end{proof}

\section{A finite adaptive advantage}\label{app:finite}

Consider the three labelled four-cycles on $\{1,2,3,4\}$, indexed by the
vertex $k\in\{1,2,3\}$ not adjacent to $4$, with two copies and the full
Pauli alphabet. The following adaptive protocol succeeds with probability
$41/48$. Measure $Z_1Z_2Z_3X_4$ on the first copy. The stabilizer of
vertex $4$ gives $x_4=z_1+z_2+z_3+z_k$, so the posterior is uniform on
$\{k:z_k=x_4+z_1+z_2+z_3\}$, whose size is $1,2,3$ with probabilities
$\frac14,\frac12,\frac14$. With one candidate we are done. With two
candidates $k,k'$, measure $X$ on $\{4,k\}$ and $Z$ elsewhere: if $k$ is
the non-neighbour of $4$ the outcome is uniform on four of sixteen
strings, otherwise on all sixteen, and maximum likelihood succeeds with
probability $7/8$. With three candidates, measuring all qubits in $X$
succeeds with probability $2/3$. In total
$\frac14+\frac12\cdot\frac78+\frac14\cdot\frac23=\frac{41}{48}$. An exhaustive
search over all pairs of Pauli settings, in exact integer arithmetic,
shows that this is optimal among adaptive protocols and that the best
nonadaptive pair of settings achieves $5/6$; $X/Z$ settings suffice for
both optima. Since the three graphs form one orbit under relabelling,
Bayes and worst-case optima coincide.

\section{Adaptive separable measurements}\label{app:separable}

Let $E_g(G)=-\log\max_{\alpha}|\langle\alpha|G\rangle|^2$, the maximum over
product states, be the geometric entanglement of $|G\rangle$.

\begin{proposition}\label{prop:separable}
For any prior on graph states and any protocol that measures each of $N$
copies separately with a separable POVM chosen from earlier classical
outcomes, the transcript $T$ satisfies $I(G;T)\le N\,\E[n-E_g(G)]$.
\end{proposition}

\begin{proof}
For a fixed separable POVM $\{E_y\}$, each effect is a positive
combination of product projectors, so $p_g(y)=\mathrm{tr}(E_y|G\rangle\langle
G|)\le2^{-E_g(g)}\mathrm{tr}E_y$. With the normalized reference law
$q(y)=\mathrm{tr}E_y/2^n$ this gives $D(p_g\|q)\le n-E_g(g)$, and for any prior
$\mu$, $I_\mu(G;Y)=\E_\mu D(p_g\|q)-D(p_\mu\|q)\le\E_\mu[n-E_g]$. Apply this
to each copy conditionally on the past and use the chain rule.
\end{proof}

This bound holds against adaptive protocols, including feed-forward
inside a copy. Proving $\E_\pi[n-E_g]=O(n\log d/d+\sqrt n)$ for the
distribution of Section~\ref{sec:regular} would give an adaptive lower
bound of $\Omega(d^2\log n/\log d)$ for $d=\Theta(\log n)$; we do not know
whether this estimate holds \citep[see][for dense random graphs]{ghh2025}.

\end{document}